\documentclass[aps,preprint,prd,nofootinbib]{revtex4-1}
\usepackage[utf8]{inputenc}
\usepackage{booktabs}
\usepackage{slashed}
\usepackage{indentfirst}
\usepackage{graphicx}
\usepackage{amsfonts,amssymb,amsmath}
\usepackage{graphicx}
\usepackage{psfrag}

\begin{document}

\title{Semi-leptonic Production of $D_{sJ}(3040)$ and $D_J(3000)$ in $B_s$ and $B$ Decays}
\author{Si-chen Li$^{[1]}$, Yue Jiang$^{[1]}$\footnote{jiangure@hit.edu.cn}, Tian-hong Wang$^{[1]}$, Qiang Li$^{[1]}$, Zhi-hui Wang$^{[2]}$,
Guo-Li Wang$^{[1]}$}
\address{$^1$Department of Physics, Harbin Institute of
Technology, Harbin, 150001\\
$^2$School of Electrical \& Information Engineering, Beifang University of Nationalities, Yinchuan, 750021}

 \baselineskip=20pt

\begin{abstract}
In this paper, we study the productions of the newly detected states $D_{sJ}(3040)$ and $D_J(3000)$ observed by BABAR Collaboration and LHCb Collaboration. We assume these states to be the $D_s(2P)$ and $D(2P)$ states with the quantum number $J^P=1^+$ in our work. The results of improved Bethe-Salpeter method indicate that the semi-leptonic decays via $B_s$ and $B$ into $D_{sJ}(3040)$ and $D_J(3000)$ have considerable branching ratios, for example, Br($\overline{B}_s^0 \rightarrow D{_{sJ}^+}(3040)e^-\overline{\nu}{_e}$)=$5.79\times10^{-4}$, Br($\overline{B}^0\rightarrow D_{J}^+(3000)e^-\overline{\nu}{_e}$)=$2.63\times10^{-4}$, which shows that these semi-leptonic decays can be accessible in experiments.

 \vspace*{0.5cm}


\end{abstract}

\maketitle

\section{INTRODUCTION}
The studys of charmed and charmed-strange mesons have made great progress in recent years, which intrigues great deal of interests in revealing their properties. More and more new resonances have been observed in experiments. For example, in charmed-strange family, $D_{s1}^*(2700)^{\pm}$ was reported by Belle Collaboration through the cascaded decay $B^+\rightarrow \overline{D}^0D_{s1}\rightarrow \overline{D}^0 D^0 K^+$ and identified as a $1^-$ assignment \cite{1}, and $D_{sJ}^*(2860)^{\pm}$ was discovered by BABAR Collaboration in $D_{sJ}(2860)\rightarrow D^0K^+, D^+K_s^0$ \cite{2}, which is very likely to be $3^-$ state. In charmed family, $D(2550)$, $D(2600)$, $D(2750)$, $D(2760)$ were observed by BaBar Collaboration with analysis of helicity distribution \cite{3}. ($D(2550)$, $D(2600)$) are tentatively identified as 2S doublet $(0^-, 1^-)$ while $D(2750)$ and $D(2760)$ are $1D$ doublet ($2^-$, $3^-$) \cite{3.1}. Recently, two new resonances have been detected experimentally with masses around $3000$ MeV, $D_{sJ}$(3040)$^+$ was observed in the $D^*K$ invariant mass spectrum in inclusive $e^+e^-$ collision by BABAR \cite{4}, which is a good candidate as the radial excitation of $D_{s1}(2460)^+$ \cite{5}. In $D^{+}\pi^{-}$ and $D^{0}\pi^{+}$ mass spectra, $D_J(3000)^0$ was observed by LHCb Collaboration \cite{6}, which could be interpreted as the radial excitation of $D_{1}(2430)^0$, and their masses and full widths are \cite{4,6}
\begin{eqnarray}
\begin{aligned}
m_{D_{sJ}(3040)^{+}}=\left(3044\pm8^{+30}_{-5}\right)\mathrm{MeV},\\
\varGamma_{D_{sJ}(3040)^{+}} =\left(239\pm35^{+46}_{-42}\right)\mathrm{MeV},\\
m_{D_{J}(3000)^{0}}=\left(2971.8\pm8.7\right)\mathrm{MeV},\\
\varGamma_{D_{J}(3000)^{0}} =\left(188.1\pm44.8\right)\mathrm{MeV}.\\
\end{aligned}
\end{eqnarray}

Regarding to the topic of radial excited states of $D_s$ and $D$ mesons, several works have been done about their mass spectra and strong decays \cite{7,7.1,7.2,7.3}. One thing drawing our attention is that no other heavy-light $2P$ state has been confirmed by experiment except charmonium and bottomonium, which means the study of charmed and charmed-strange $2P$ states will enlarge our knowledge of bound states and deepen the understanding of nonperturbative QCD.

We notice that $D_{sJ}(3040)$ and $D_J(3000)$, assumed to be radial excitation of $D_{s1}(2460)$ and $D_{1}(2430)$ in recent studies, can be produced via the semileptonic decays of  $B_{s}$ and $B$, which are different from the observed production processes. Previous studies show that semi-leptonic decays could be a good platform to produce charmed and charmed-strange mesons, for instance, the process of $B_{s}\rightarrow D_{s1}(2460)l\overline\nu_{l}$ has been calculated through relativistic quark model based on the quasipotential approach \cite{7.5}, three point QCD sum rule methods \cite{7.6}, QCD sum rules under HQET \cite{7.7}, constituent quark meson model \cite{7.8}, and instantaneous Bethe-Salpeter method \cite{8}. The same order $10^{-3}$ of the results in various models indicates that semi-leptonic decays have considerable branching ratios. In addition, the study of semi-leptonic decay provide an extra source of information for the determination of CKM matrix elements and the relativistic quark dynamics inside heavy-light mesons. In this paper, we explore the production of $D_{sJ}(3040)$ and $D_J(3000)$ by the improved B-S(Bethe-Salpeter) method, and give the results of form factors as well as branching ratios.

The rest of this paper is organized in the following arrangements. In section 2 we deduce the formulation of semi-leptonic decay. The hadronic matrix elements of production are given in section 3, numerical results and discussions are presented in section 4.

\section{THE FORMULATIONS OF SEMI-LEPTONIC DECAY}

We take ${\overline{B}^{0}_{s}}\rightarrow D^{+}_{sJ}(3040) l^{-}_{}\overline\nu_{l}$ as an example to illustrate this type of process.
The feynman diagram of this semi-leptonic decay is drawn in figure 1.
\begin{figure}[!hbt]
\centering
\includegraphics[height = 5cm, width = 9 cm]{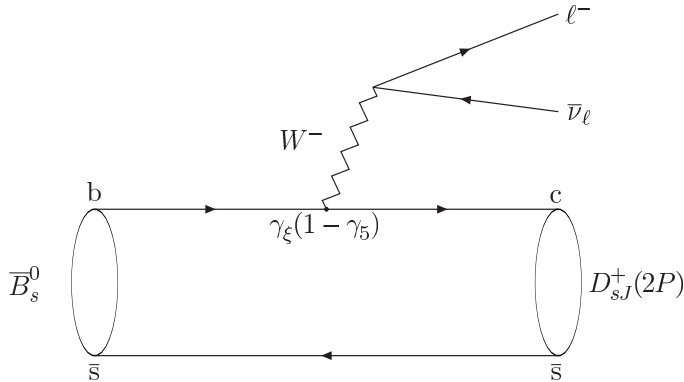}
\caption{Feynman diagram of semi-leptonic decay ${\overline{B}^{0}_{s}}\rightarrow D^{+}_{s}(2P) l^{-}_{}\nu_{l}$ }
\label{fig:XXXX} 
\end{figure}

The amplitude of ${\overline{B}^{0}_{s}}\rightarrow D^{+}_{sJ}(3040) l^{-}_{}\overline\nu_{l}$ is \cite{8}

\begin{equation}
  T=\frac{G_F}{\sqrt{2}}V_{cb}\overline{u}(p_l)\gamma ^{\xi }(1-\gamma _5)\nu (p_{\nu _{l}})\left\langle D_{sJ}^+(3040)(P_f) |J_\xi  |  \overline{B}_{s}^0(P)\right\rangle ,
\end{equation}
where $V_{cb}$ is the CKM matrix element, $G_F$ is the fermi constant, $J_\xi =V_\xi -A_\xi $ is  the charged weak current, in which $V_\xi =\overline{c}\gamma_\xi b $, $A_\xi=\overline{c} \gamma_\xi\gamma_5b$, $P$ and $P_f$ are the momenta of the initial meson $\overline{B}_{s}^0$ and final meson $D_{sJ}^+(3040)$ respectively. Thus the square of the amplitude is:\\
\begin{equation}
|T|^2=\frac{G_F^2}{2}|V_{bc}|^2l^{\xi\xi'} h_{\xi\xi'},
\end{equation}
where the leptonic tensor could be simplified as:
\begin{equation}
l^{\xi\xi'}=8\left(p_{\nu{_l}}^{\xi} p_{l}^{\xi' } +p_{l}^{\xi }p_{v_{l}}^{\xi'}-p_{v_l}p_{l}g^{\xi \xi' }+i\epsilon^{\xi \xi' \alpha \beta }p_{1\alpha }p_{2\beta } \right),
\end{equation}
and hadronic tensor is defined as:
\begin{equation}
h_{\xi\xi'}=\left\langle \overline{B}_{s}^0(P) |J _\xi ^{ \dagger} | D_{sJ}^+(3040)(P_f)\right\rangle  \left\langle D_{sJ}^+(3040)(P_f) |J_{\xi'} |  \overline{B}_{s}^0(P)\right\rangle,
\end{equation}
which can be described as form factors. Explicit forms are present in next subsection.


\section{HADRONIC MATRIX ELEMENT OF SEMI-LEPTONIC DECAY}

The calculation of hadronic matrix element is model-dependent. In this paper, we determine the hadronic matrix element through the instantaneous Bethe-Salpeter method with Mandelstam formalism. As a relativistic quark model, the instantaneous Bethe-Salpeter method has been applied in many transitions among heavy-light mesons. More details about instantaneous Bethe-Salpeter equation are given in Appendix A.

Regarding to the classification of heavy-light meson, the heavy-light mesons can be classified in doublets based on the total angular momentum of the light quark $s_l$. We can categorize the heavy mesons into several doublets, for example, the S doublet is $(0^+, 1^+)$ with $s_l = \frac{1}{2}$ , and the T doublet is $(1^+, 2^+)$ with $s_l = \frac{3}{2}$ , thus the $1^+$ states can be labeled as $P_1^{1/2}$ and $P_1^{3/2}$. But in our method, we solved the Salpeter equation and obtained the wave functions of the $^3P_1$ and $^1P_1$ states, whose forms are given in Appendix B, then the physical states are mixtures of the $^3P_1$ and $^1P_1$:
\begin{equation} \label{mixing}
\begin{aligned}
  &\left|\frac{3}{2} \right \rangle = {\cos{\theta}}\left|^1P_1\right\rangle+ {\sin {\theta}}\left |^3P_1\right\rangle ,\\
  &\left|\frac{1}{2} \right \rangle = -{\sin{\theta}}\left|^1P_1\right\rangle +{\cos {\theta}} \left|^3P_1\right\rangle.
\end{aligned}
\end{equation}

In the heavy quark limit, which is $m_Q\rightarrow \infty $, the mixing angle $\theta\approx 35.3^{\circ}$ \cite{9}. $D_{sJ}(3040)$ is assumed to be the radial excitation of  $D_{s1}(2460)$ in this paper, which is $P_1^{1/2}$ state. The partner has not been discovered yet, which is correspondent to $P_1^{3/2}$ state. By the B-S method with the instantaneous approach, the hadronic matrix element can be written as the overlapping integral over the initial and final B-S wave functions \cite{8}:
\begin{eqnarray}
\nonumber \left\langle  D_{sJ}^+\left(P_f\right) \left(^1{ P}_1\right)|J_{\xi}  | \overline{B}_s^0\left(P\right) \right\rangle&&=i \int\frac{{\rm d}^4q}{(2\pi )^4} {\rm{Tr}} \left[ \overline{\chi}_{D_{sJ}}(P_f, P, q_1)(\alpha_1 \slashed{P} + \slashed{q}-m_s ) \gamma_\xi (1-\gamma_5) \chi_{B_s^0}(P, q)    \right]\\
\nonumber&&=\int\frac{{\rm d}\vec{q}}{(2\pi)^3}{\rm{Tr}}\left[ \overline{\varphi}_{1^{+}}^{++}\left(^1{ P}_1\right)\left(\vec{q}_1\right)\gamma_\xi  \left(1-\gamma_5\right) {\varphi}_{0^{-}}^{++}\left(\vec{q}\right) \frac{ \slashed{P}}{M}\right]\\
&&=\epsilon_\mu \left(t_1P_\xi  P^\mu +t_2P_{f\xi} P^\mu +t_3g^{\; \;\mu} _{ \xi} +t_4\epsilon ^{\; \; PP_1\mu }_\xi\right),
\end{eqnarray}
\begin{eqnarray}\hspace{-0.7in}
\nonumber \left\langle  D_{sJ}^+(P_f) \left(^3{ P}_1\right)|J_{\xi}| \overline{B}_s^0(P) \right\rangle&&=\int\frac{{\rm d}\vec{q}}{(2\pi)^3}{\rm{Tr}}\left[\overline{\varphi}_{1^{+}}^{++}\left(^3{ P}_1\right)\left(\vec{q}_1\right)\gamma_\xi  \left(1-\gamma_5\right) {\varphi}_{0^{-}}^{++}(\vec{q}) \frac{ \slashed{P}}{M}\right] \\
&&=\epsilon_\mu \left(t_5P_\xi  P^\mu +t_6P_{f\xi} P^\mu +t_7g^{\; \;\mu} _\xi +t_8\epsilon ^{\; \; PP_1\mu }_\xi\right),
\end{eqnarray}
where $\vec{q}$ and $\vec{q}_1$ are relative three-momentum between the quark and anti-quark for initial state and final state. $t_1$ to $t_8$ are the form factors, which are given in Appendix C.

The wave functions we adopt above are for $^1P_1$ and $^3P_1$ states. Due to the mixture of physical states, the form factors for $P^{1/2}$ and $P^{3/2}$ states are given as:
\begin{eqnarray}
\begin{aligned}
x_{i+4}= t_i\cos\theta +t_{i+4}\sin\theta ,\\
x_i=-t_i\sin\theta +t_{i+4}\cos\theta ,
\end{aligned}
\end{eqnarray}
where $i=1, 2, 3, 4.$

Another thing we should notice is that the masses of $^1P_1$ and $^3P_1$ are different from $P^{1/2}$ and $P^{3/2}$. There is also a mixture between them and the relation is given as \cite{11}:
\begin{eqnarray}
\begin{aligned}
&m^2_{^1P_1}=m^2_{\frac{1}{2}} \sin^2\theta +m^2_{\frac{3}{2}} \cos ^2 \theta,\\
&m^2_{^3P_1}=m^2_{\frac{1}{2}} \cos^2\theta +m^2_{\frac{3}{2}} \sin ^2 \theta.
\end{aligned}
\end{eqnarray}

By giving the form factors, the width of semi-leptonic decay is
\begin{eqnarray}
\begin{aligned}
\varGamma  =&\frac{G_F^2V_{cb}^2M^3}{32\pi ^3} \int \frac{p_l}{E_l} {\rm d} \vec{p}_l \int \frac{p_f}{E_f}{\rm d}\vec{p}_f \left \{ 2\alpha \left( \frac{y}{M^2}\right ) +\beta_{++} \left[ 4\left (2x \left (1-\frac{M_f^2}{M^2}+y\right )-4x^2-y\right )  \right. \right.  \\
&\left. \left. +\frac{m_l^2}{M^2}\left (8x+4\frac{M_f^2}{M^2}-3y-\frac{m_l^2}{M^2}\right )\right]+\left(\beta_{\pm }+\beta_{\mp}\right)\frac{m_l^2}{M^2}\left(2-4x+y-2\frac{M_f^2}{M^2}\right)\right.\\
&\left.+ \beta_{--} \frac{m_l^2}{M^2}\left(y-\frac{m_l^2}{M^2}\right)
+ 2\gamma \left[y\left(1-4x+y-\frac{M_f^2}{M^2}\right)+ \frac{M_l^2}{M^2}\left(1+y-\frac{M_f^2}{M^2}\right) \right]  \right \},
\end{aligned}
\end{eqnarray}
where $M_f$ and $M$ are masses of the final and initial meson respectively, $m_l$ is the mass of the corresponding lepton. $\alpha$, $\beta_{\pm \pm}$ and $\gamma$ are coefficients as functions of the form factors:

\begin{equation}
\begin{aligned}
&x=\frac{E_l}{M},  y=\frac{\left(p-p_f\right)^2}{M^2},\\
&\alpha =x_3^2+x_4^2  M^2 p_f^2,\\
&\beta_{++} =p_f^2 \frac{\left(x_1+x_2\right)^2}{4 M_f^2}+\frac{\left(2ME_f-M^2-M_f^2\right) x_4^2}{4}+\frac{x_3^2}{4 M_f^2}+\left(\frac{ME}{M_f^2}-1\right)\frac{\left(x_1+x_2\right)x_3}{2M},\\
&\beta_{+-} =\beta_{-+}=p_f^2\frac{(x_1+x_2)(x_1-x_2)}{4M_f^2}+\frac{\left(M^2-M_f^2\right)}{4}-\left(x_1+\frac{x_2EM}{M_f^2}\right)\frac{x_3}{2M}-\frac{x_3^2}{4M_f^2},\\
&\beta_{--} =p_f^2\frac{\left(x_1-x_2\right)^2}{4M_f^2}-\frac{\left(2ME+M_f^2+M^2\right)x_4^2}{4}+\frac{x_3^2}{4M_f^2}+\left(1+\frac{ME}{M_f^2}\right)\frac{\left(x_2-x_1\right)x_3}{2M},\\
&\gamma=-x_3 x_4.
\end{aligned}
\end{equation}
\section{NUMERICAL RESULTS AND ANALYSIS}

\subsection{form factors}
In our model, the input parameters of calculation are chosen as following: $\lambda$ =0.21 GeV$^2$, $\varLambda _{\rm QCD}$=0.27 GeV, a=e=2.71, $\alpha $=0.06 GeV, $m_b$=4.96 GeV, $m_s$=0.50 GeV, $m_c$=1.62 GeV, $m_d$=0.311 GeV, which are the best results to fit the mass spectrum of related mesons \cite{12}. For semi-leptonic decay, we also need CKM matrix elements: $V_{bc}$=0.0406, and the lifetime of initial meson $\tau _{B_{s0}}=1.469\times 10^{-12}$ s, the masses of $m_{B^0}$=5279.58 MeV and $m_{B^0_s}$=5366.77 MeV are taken from PDG \cite{13}. We notice that the partners of $D_{sJ}(3040)$ and $D_J(3000)$ are not discovered yet, the masses required in our calculation are taken as 3022.3 MeV and 2913.8 MeV for $D_{s}(2P^{3/2}_{1})$ and $D(2P^{3/2}_{1})$ respectively. Varying all the input parameters simultaneously within $\pm$ 5\% of the central values, we obtain the uncertainties of branching ratios.

\begin{figure}[htbp]
\centering
\includegraphics[height = 6cm, width = 8 cm]{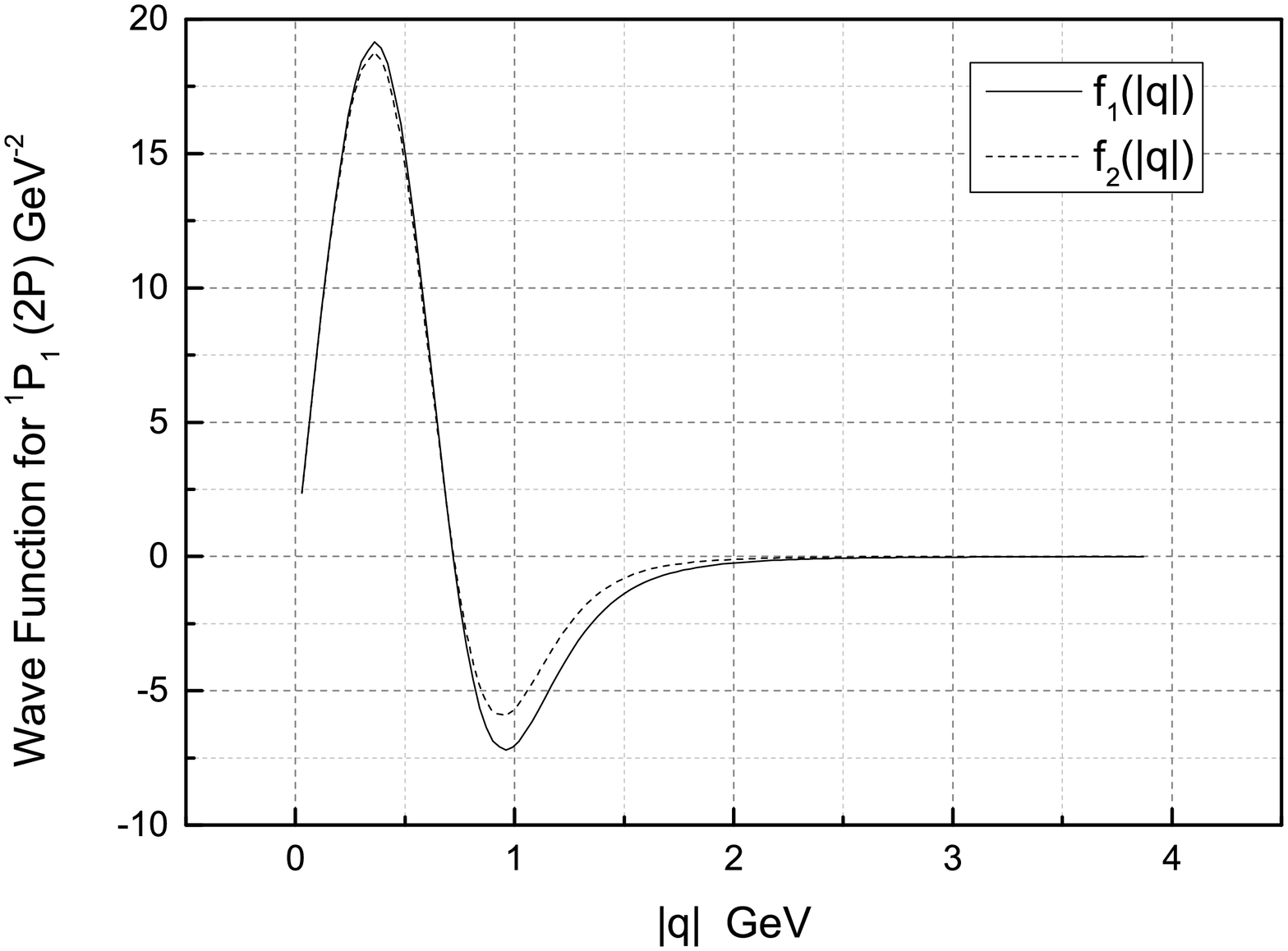}
\includegraphics[height = 6cm, width = 8 cm]{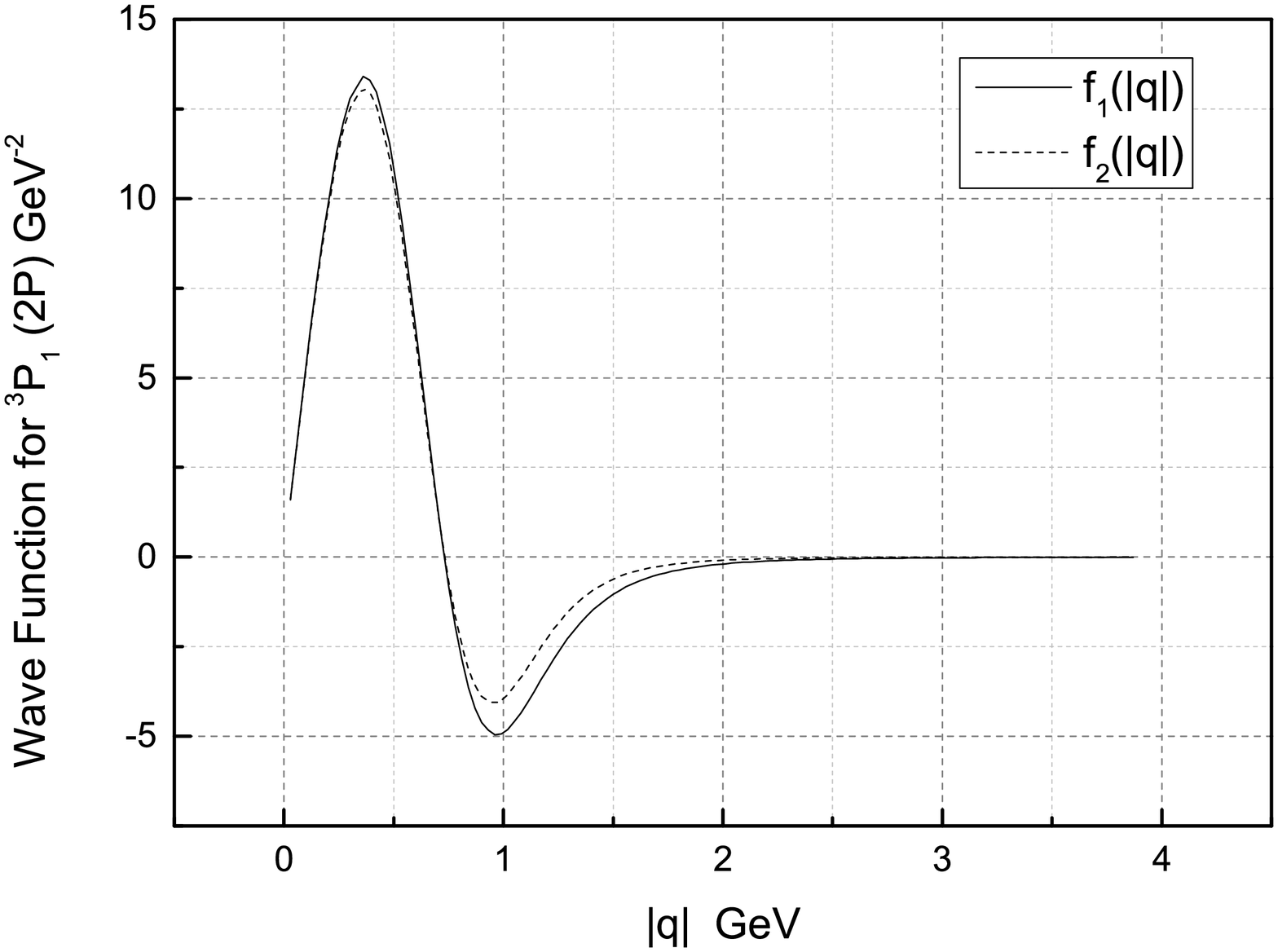}
\caption{The wavefunctions of $^1P_1$ and $^3P_1$ for $D_{s}(2P)$ meson }
\label{fig:XXXX} 
\end{figure}

 To show the numerical results of wave functions explicitly, we plot the $^1P_1$ and $^3P_1$ state for $D_s(2P)$ meson in figure 2. We can see that $^1P_1$ and $^3P_1$ states share the same shape. As an example, The form factors $x_1$ to $x_4$ are shown in figure 3, where $t=(P-P_f)^2=M^2+M_f^2-2ME_f$ and $t_m$ is the maximum of $t$.
\begin{figure}[!hbt]
\centering
\includegraphics[height = 7cm]{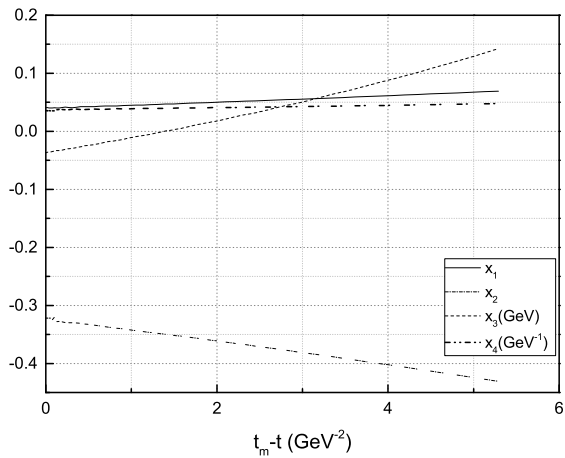}
\caption{The form factors of $\overline{B}^{0}_{s}\rightarrow D_{sJ}(3040)^{+} e^{-}\overline\nu_{e}$}
\label{fig:XXXX} 
\end{figure}
\subsection{branching ratios}
\subsection*{for $D_{sJ}(3040)$}
In table \uppercase\expandafter{\romannumeral1}, we show the branching ratios of semi-leptonic production of $D_{s}(2P)^{+}$. Generally, the cases of $e$ and $\mu$ are 2 orders of magnitude larger than the case of $\tau$ due to the phase space. We also notice that the branching ratios of $\overline{B}_s^0\rightarrow D{_{sJ}^+}(P{_{1}^{3/2}})l^-\overline{\nu}{_l}$ are 10 times larger than $\overline{B}_s^0\rightarrow D{_{sJ}^+}(P{_{1}^{1/2}})l^-\overline{\nu}{_l}$. Ref \cite{14} calculate the same process via covariant light-front quark model. The result in Ref \cite {15} is obtained through modified harmonic-oscillator light-front wave function (\uppercase\expandafter{\romannumeral1}) and light-front quark model associated within HQET (\uppercase\expandafter{\romannumeral2}). We can see that our results are well consistent with the light-front quark model associated within HQET but show a little discrepancy with the other two results. All these results indicate that more theoretical researches should be done in the future.

\begin{center}
\begin{table*}[h]
\caption{\small Branching ratios of ${\overline{B}^{0}_{s}}\rightarrow D^{+}_{s}(2P) l^{-}_{}\overline\nu_{l}$}
\begin{center}
\begin{tabular}{cccccc}
\hline\hline
\rule[-2mm]{0mm}{6.5mm}{}&{\bfseries ours}&{\bfseries \cite{14}}&{\bfseries \uppercase\expandafter{\romannumeral1} \cite{15}}&{\bfseries \uppercase\expandafter{\romannumeral2} \cite{15}}\\
\hline
\rule[-2mm]{0mm}{6.5mm}
$\overline{B}_s^0\rightarrow D{_{sJ}^+}(3040)e^-\overline{\nu}{_e}$ &  $(5.79^{+2.1}_{-2.0})\times 10^{-4}$ & ${}$ & $(2.49^{+0.4}_{-0.4})\times 10^{-4}$& ${5.6\times 10^{-4}}$\\
\rule[-2mm]{0mm}{6.5mm}
$\overline{B}_s^0\rightarrow D{_{sJ}^+}(P{_{1}^{3/2}})e^-\overline{\nu}{_e}$&$(2.34^{+1.30}_{-1.04})\times 10^{-3}$&${}$&$(2.42^{+0.07}_{-0.14})\times 10^{-3}$& ${1.24\times 10^{-3}}$\\
\rule[-2mm]{0mm}{6.5mm}
$\overline{B}_s^0\rightarrow D{_{sJ}^+}(3040)\mu^- \overline{\nu}{_\mu }$&$(5.77^{+2.15}_{-2.07})\times 10^{-4}$&$(3.5^{+1.1}_{-1.0})\times 10^{-4}$&$(2.46^{+0.4}_{-0.42})\times 10^{-4}$& ${5.6\times 10^{-4}}$\\
\rule[-2mm]{0mm}{6.5mm}
$\overline{B}_s^0\rightarrow D{_{sJ}^+}(P{_{1}^{3/2}})\mu^- \overline{\nu}{_\mu }$&$(2.36^{+1.28}_{-1.06})\times 10^{-3}$&$(4.0^{+0.4}_{-0.5})\times 10^{-3}$&$(2.39^{+0.07}_{-0.13})\times 10^{-3}$& ${1.24\times 10^{-3}}$\\
\rule[-2mm]{0mm}{6.5mm}
$\overline{B}_s^0\rightarrow D{_{sJ}^+}(3040)\tau^- \overline{\nu}{_\tau }$&$(4.07^{+1.95}_{-1.74})\times 10^{-6}$&$(9.9^{+4.4}_{-3.5})\times 10^{-6}$&$(5.2^{+0.4}_{-0.5})\times 10^{-6}$\\
\rule[-2mm]{0mm}{6.5mm}
$\overline{B}_s^0\rightarrow D{_{sJ}^+}(P{_{1}^{3/2}})\tau^- \overline{\nu}{_\tau }$&$(3.49^{+2.39}_{-1.78})\times 10^{-5}$&$(9.7^{+0.8}_{-0.8})\times 10^{-5}$&$(0.43^{+0}_{-0.01})\times 10^{-6}$\\
\hline\hline
\end{tabular}
\end{center}
\end{table*}
\end{center}

Due to the lack of data of $D_s$(2$P$) state, as a comparison, we give the information about $1P$ state with $J^P=1^+$. The branching ratio of cascaded decay Br$(B_s^0\rightarrow D_{s1}(2536)^-\mu ^+\nu _{\mu })\times $Br$(D_{s1}(2536)^-\rightarrow D^{*-}K^0_s$ )=$(2.5\pm0.7)\times 10^{-3}$, and the branching ratio of strong decay is $0.85\pm0.12$ \cite{13}, so the branching ratio of semi-leptonic decay into $1P$ state is $2.94^{+1.44}_{-1.09}\times 10^{-3}$. The corresponding first radial excitation of $D_{s1}(2536)^-$ is $D_{s1}(P_1^{3/2})^-$, whose production rate via semi-leptonic decay is $2.34\times 10^{-3}$ in our method \cite{8}, this may imply that our results are reliable.

Although the production ratio of $D_{sJ}(3040)$ is very small in $\overline{B}^0_{s}$ semi-leptonic decay, considering that the LHCb experiment will produce more than $10^6$ $B_s$ mesons per running year \cite{15}, the branching ratios of $\overline{B}^0_s\rightarrow D_{sJ}(3040)^{+} e^{-}\overline\nu_{e}$ around $10^{-4}$ are considerable, and are accessible in the current $B_s$ decay data. So the semi-leptonic approach has a promising prospect in producing $D_{sJ}(3040)$.
\subsection*{for $D_J(3000)$}

\begin{center}
\begin{table*}[h]
\caption{\small Branching ratios of $\overline{B}^0\rightarrow D^{+}(2P) l^{-}_{}\overline\nu_{l}$}
\begin{center}
\begin{tabular}{cccccc}
\hline\hline
\rule[-2mm]{0mm}{6.5mm}{}&{\bfseries ours}&{\bfseries \cite{15}}\\
\hline
\rule[-2mm]{0mm}{6.5mm}
$\overline{B}^0\rightarrow D_J(3000)^+ e^-\overline{\nu}{_e}$ &  $(2.63^{+0.33}_{-0.68})\times 10^{-4}$ & $({2.57^{+0.39}_{-0.44})\times 10^{-4}}$ \\
\rule[-2mm]{0mm}{6.5mm}
$\overline{B}^0\rightarrow D(2P{_{1}^{3/2}}){^+}e^-\overline{\nu}{_e}$&$(2.62^{+0.64}_{-0.50})\times 10^{-4}$&$({2.72^{+0.02}_{-0.11})\times 10^{-3}}$\\
\rule[-2mm]{0mm}{6.5mm}
$\overline{B}^0\rightarrow D_J(3000)^+\mu^- \overline{\nu}{_\mu }$&$(2.38^{+0.60}_{-0.42})\times 10^{-4}$&$(2.54^{+0.38}_{-0.44})\times 10^{-4}$\\
\rule[-2mm]{0mm}{6.5mm}
$\overline{B}^0\rightarrow D(2P{_{1}^{3/2}}){^+}\mu^- \overline{\nu}{_\mu }$&$(2.42^{+0.57}_{-0.46})\times 10^{-4}$&$(2.69^{+0.02}_{-0.11})\times 10^{-3}$\\
\rule[-2mm]{0mm}{6.5mm}
$\overline{B}^0\rightarrow D_J(3000)^+ \tau^- \overline{\nu}{_\tau }$&$(1.81^{+0.54}_{-0.30})\times 10^{-6}$&$(5.2^{+0.4}_{-0.5})\times 10^{-6}$\\
\rule[-2mm]{0mm}{6.5mm}
$\overline{B}^0\rightarrow D(2P{_{1}^{3/2}}){^+}\tau^- \overline{\nu}{_\tau }$&$(4.44^{+0.76}_{-0.59})\times 10^{-6}$&$(0.603^{+0}_{-0.02})\times 10^{-4}$\\
 \hline\hline
\end{tabular}
\end{center}
\end{table*}
\end{center}

In table \uppercase\expandafter{\romannumeral2}, the results of $\overline{B}^0\rightarrow D^{+}(2P) l^{-}_{}\overline\nu_{l}$ are presented. Our results show that the branching ratios into two doublets are of the same order of $ 10^{-4}$ for $e$ and $\mu$, $10^{-6}$ for $\tau$. While the results from light front quark model \cite{15} are the same of $10^{-4}$ for $2P^{1/2}_{1}$ state, but one order of magnitude smaller than ours for $2P^{3/2}_{1}$ state. To give some clues for this discrepancy, we list the results of $\overline{B}^0\rightarrow D^{+}(1P) l^{-}_{}\overline\nu_{l}$ as the comparison. In table \uppercase\expandafter{\romannumeral3}, we give the cascaded decay of $D(1P)$ states, in which the  $D_1(2430)$ and  $D_1(2420)$ are $D(1P_1^{1/2})$ and $D(1P_1^{3/2})$ respectively.

Considering that the strong decays of $D_1$ state are dominant channels at around $67\%$ due to the isospin symmetry, one thing we should notice in table \uppercase\expandafter{\romannumeral3} is that for $D_1(1P^{3/2}_1$) and $D_1(1P^{1/2}_1$), the branching ratios of semi-leptonic productions are almost the same of $4.5\times 10^{-3}$ in experiment. Our results are consistent with this data. If the behaviors of $2P$ states are similar to $1P$ states, our results seem to be more reasonable.
\begin{center}
\begin{table*}[h]
\caption{\small Cascaded decay of $\overline{B}^0$ into $D^-(1P)$}
\begin{center}
\begin{tabular}{cccccc}
\hline\hline
\rule[-2mm]{0mm}{6.5mm}{}&{\bfseries ours}&{\bfseries exp\cite{13}}\\
\hline
\rule[-2mm]{0mm}{6.5mm}
${\rm Br}(\overline{B}^0\rightarrow D_1(2430)^- l^+\overline{\nu}{_l})\times {\rm Br}(D_1(2430)^- \rightarrow \overline{D}^{*0}\pi^{-})$ &  $3.92^{+0.30}_{-0.39}\times 10^{-3}$ & ${(3.1\pm 0.9)\times 10^{-3}}$ \\
\rule[-2mm]{0mm}{6.5mm}
${\rm Br}(\overline{B}^0\rightarrow D_1(2420)^- l^+\overline{\nu}{_l})\times {\rm Br}( D_1(2420)^-\rightarrow \overline{D}^{*0}\pi^{-})$&$5.51^{+0.07}_{-0.14}\times 10^{-3}$&${(2.80\pm0.28)\times 10^{-3}}$\\
 \hline\hline
\end{tabular}
\end{center}
\end{table*}
\end{center}

Similar with $B^0_s\rightarrow D^{+}_{s}(2P) l^{-}_{}\overline\nu_{l}$, the branching ratios are large enough to be observed in experiment, so we suggest that the LHCb and Belle \uppercase\expandafter{\romannumeral2} Collaboration carry out the study of semi-leptonic decays above.

The possible sources of the uncertainty on the results may come from these following factors: (1) The spin partners of $D_J(3000)$ and $D_{sJ}(3040)$ are not detected experimentally yet. In our work, the masses of $D(2P^{3/2}_{1})$ and $D_{s}(2P^{3/2}_{1})$ are assumed to be around 3000 MeV and 2913 MeV. It is one of the important sources of uncertainty. (2) $P^{1/2}$ and $P^{3/2}$ states are mixture of $^1P_1$ and $^3P_1$ states. The mixing equation we use in this paper is determined by the mixing angle, and this angle we use is derived from heavy-quark limit, which deviates from the realistic mixing angle, especially for the higher radial excitations \cite{16}. That is another possible way for the uncertainty to be increased. These sources show that there are a lot of researches to be done in the future to reduce the uncertainty and make the prediction more precise.


\subsection*{for $3P$ states}

Although no $3P$ state of $D_s$ or $D$ meson has been observed in experiment yet, we give a very preliminary prediction in our method. The masses we used are 3421 MeV and 3427 MeV for $D_s(3^1P_1)$ and $D_s(3^3P_1)$ states, 3215 MeV and 3220 MeV for $D(3^1P_1)$ and $D(3^3P_1)$ states, which are predicted in our model. The mixing angles $\theta\approx 35.3^{\circ}$. The results are given in table \uppercase\expandafter{\romannumeral4}.

\begin{center}
\begin{table*}[h]
\caption{\small Branching ratios of $3P$ states of $D_s$ and $D$ meson}
\begin{center}
\begin{tabular}{cccccc}
\hline\hline
\rule[-2mm]{0mm}{6.5mm}{}&{\bfseries Br}&{}&{}&{\bfseries Br}\\
\hline
\rule[-2mm]{0mm}{6.5mm}
$\overline{B}_s^0\rightarrow D_s(3P{_{1}^{1/2}})^+ e^-\overline{\nu}{_e}$ &  $(7.24^{+2.65}_{-2.18})\times 10^{-6}$ &{\quad}& $\overline{B}^0\rightarrow D(3P{_{1}^{1/2}})^+ e^-\overline{\nu}{_e}$& $(2.35^{+0.29}_{-0.28})\times 10^{-6}$\\
\rule[-2mm]{0mm}{6.5mm}
$\overline{B}_s^0\rightarrow D_s(3P{_{1}^{3/2}}){^+}e^-\overline{\nu}{_e}$&$(2.70^{+0.40}_{-0.31})\times 10^{-4}$ & {\quad}& $\overline{B}^0\rightarrow D(3P{_{1}^{3/2}}){^+}e^-\overline{\nu}{_e}$&$(3.48^{+0.15}_{-0.12})\times 10^{-4}$\\
\rule[-2mm]{0mm}{6.5mm}
$\overline{B}_s^0\rightarrow D_s(3P{_{1}^{1/2}})^+\mu^- \overline{\nu}{_\mu }$&$(7.32^{+2.69}_{-2.21})\times 10^{-6}$ & {\quad}& $\overline{B}^0\rightarrow D(3P{_{1}^{1/2}})^+\mu^- \overline{\nu}{_\mu }$&$(2.36^{+0.29}_{-0.28})\times 10^{-6}$\\
\rule[-2mm]{0mm}{6.5mm}
$\overline{B}_s^0\rightarrow D_s(3P{_{1}^{3/2}}){^+}\mu^- \overline{\nu}{_\mu }$&$(2.68^{+0.40}_{-0.31})\times 10^{-4}$ & {\quad}& $\overline{B}^0\rightarrow D(3P{_{1}^{3/2}}){^+}\mu^- \overline{\nu}{_\mu }$&$(3.47^{+0.14}_{-0.12})\times 10^{-4}$\\
\rule[-2mm]{0mm}{6.5mm}
$\overline{B}_s^0\rightarrow D_s(3P{_{1}^{1/2}})^+ \tau^- \overline{\nu}{_\tau }$&$(7.36^{+2.33}_{-2.09})\times 10^{-10}$ &{\quad}& $\overline{B}^0\rightarrow D(3P{_{1}^{1/2}})^+ \tau^- \overline{\nu}{_\tau }$ &$(7.35^{+0.85}_{-0.87})\times 10^{-9}$ \\
\rule[-2mm]{0mm}{6.5mm}
$\overline{B}_s^0\rightarrow D_s(3P{_{1}^{3/2}}){^+}\tau^- \overline{\nu}{_\tau }$&$(1.62^{+0.18}_{-0.14})\times 10^{-7}$ &{\quad}& $\overline{B}^0\rightarrow D(3P{_{1}^{3/2}}){^+}\tau^- \overline{\nu}{_\tau }$ &$(1.17^{+0.06}_{-0.05})\times 10^{-6}$\\
 \hline\hline
\end{tabular}
\end{center}
\end{table*}
\end{center}

In table \uppercase\expandafter{\romannumeral4}, the branching ratios of $3P$ states are much lower than those of $2P$ states, which presents challenges in current experiment. In addition, we see an interesting result that two mixing $3P$ states of $D$ meson show discrepancy in semi-leptonic decay of $\overline{B}^0$, which needs more data and researches to give a more precise result.

\section{SUMMARY}
The accumulative data of charmed and charmed-strange mesons are becoming more and more abundant with the running of colliders. The study of higher radial excitation in charmed and charmed-strange families is becoming a intriguing field. Two of the newly detected states are $D_{sJ}(3040)^+$ and $D_J(3000)^0$, which are very likely to be $D_{s}(2P)$ and $D(2P)$ states. The productions of these states in experiment are the inclusive $e^+e^-$ interaction and $D\pi$ channel.

Under the instantaneous Bethe-Salpeter framework, we have studied the branching ratios of semi-leptonic decays into $D_{sJ}(3040)$ and $D_J(3000)$. Our results indicate that the semileptonic production from $B_s$ and $B$ can be a good platform to produce considerable amount of $D_{sJ}(3040)$ and $D_J(3000)$, so we urge that relevant experiment groups could focus on these channels. Those phenomenological investigations are important to further experimentally study of $2P$ state of $D_s$ and $D$ meson.


 \section*{Acknowledgements}
This work was supported in part by the National Natural Science Foundation of China (NSFC)
under Grant Nos. 11505039, 11575048, 11405004 and 11405037, and in part by PIRS of HIT Nos. Q201504, B201506, A201409, and T201405.


\section*{Appendix A. Instantaneous Bethe-Salpeter equation}

We define the B-S wavefunction as:
\begin {equation} \renewcommand\theequation{A.1}
\chi _P(q)=\int d^4 x \; exp(iq\cdot x)\left\langle 0 | T[\psi_1(\alpha_2x) \overline{\psi}_2(-\alpha _1x) ] |P , \beta  \right\rangle,
\end{equation}
where $\chi_P(q)$ is the B-S wavefunction of the relevant bound state. $\beta$ is the index other than momentum, $\alpha_1=\frac{m_1}{m_1+m_2}$, $\alpha_2=\frac{m_2}{m_1+m_2}$, $q=\alpha_2 p_1 - \alpha_1 p_2$, $p_1$, $p_2$ and $m_1$, $m_2$ are the momenta and constituent masses of the quark and anti-quark, respectively. $P$ is the momentum of the initial state while $\beta$ is the quantum index to identify the state other than momentum. $q_P$ denotes $\frac{q\cdot P}{\sqrt{P^2}}$ and $q_\perp=q_{P_\perp}=q-\frac{q\cdot P}{P^2}P$.

The B-S equation in momentum space can be written as:

\begin {equation}\renewcommand\theequation{A.2}
\left(\slashed{p}_1-m_1\right)\chi _P(q)\left(\slashed{p}_2+m_2\right)=i\int \frac{d^4k}{(2\pi)^4}V(P, k, q)\chi _P(k).
\end{equation}

In the instantaneous approximation, the integral kernel takes a simple form:

\begin {equation}\renewcommand\theequation{A.3}
V(P, k, q)=V(|k-q|).
\end{equation}

Three-dimensional wavefunction can be written as:
\begin {equation}\renewcommand\theequation{A.4}
\varphi(q^\mu _{p{_\perp}}) =i\int \frac{dq_p}{2\pi}\chi _P(q).
\end{equation}

Thus, the B-S equation can be rewritten as:
\begin {equation}\renewcommand\theequation{A.5}
\chi _P(q)=S_1(p_1)\eta (q_{P_{\perp}})S_2(p_2),
\end{equation}
where
\begin{equation*}
\eta (q_{P^\mu _{\perp}})=\int\frac{d^3k_{P_\perp}}{(2\pi )^3}V(k^\mu _{P_\perp}, q^\mu _{P_\perp})\varphi (k^\mu _{P_\perp}).
\end{equation*}

The full Salpeter equation takes the form:
\begin {equation}\renewcommand\theequation{A.6}
\begin{aligned}
\left(M-\omega _{1p}-\omega _{2p}\right)\varphi ^{++}(q_{P_\perp})&=\Lambda^+_1(P_{1p_{\perp}}) \eta (q_{P_{\perp}}) \Lambda_2^+ (P_{2p_{\perp}}),\\
(M+\omega _{1p}+\omega _{2p})\varphi ^{--}(q_{P_\perp})&=-\Lambda^-_1(P_{1p_{\perp}}) \eta (q_{P_{\perp}}) \Lambda^-_2 (P_{2p_{\perp}}),\\
\varphi ^{+-}(q_{P_\perp})&=0,\\
\varphi ^{-+}(q_{P_\perp})&=0,\\
\varphi ^{\pm \pm}(q_{P_\perp})&=\Lambda^{\pm}_1(q_{P_\perp}) \frac{\slashed{P}}{M }\varphi(q_{P_\perp}) \frac{\slashed{P}}{M} \Lambda^{\pm}_2(q_{P_\perp}).
\end{aligned}
\end{equation}

In order to do the numerical integral, we need the explicit form of integral kernel. In this work, we choose the Cornell potential, which was widely used in this interaction. The Cornell potential is the sum of a linear scalar interaction and a vector interaction.

\begin {equation}\renewcommand\theequation{A.7}
\begin{aligned}
V(q)&=V_s(q)+V_v(q)\gamma ^0\otimes \gamma _0,\\
V_s(q)&=-\left(\frac{\lambda }{\alpha }+V_0\right)\delta ^3(q)+\frac{\lambda }{\pi ^2}\frac{1}{(q^2+\alpha ^2)^2},\\
V_v(q)&=-\frac{2}{3\pi ^2}\frac{\alpha _s(q)}{q^2+\alpha ^2},\\
\alpha _s(q)&=\frac{12\pi}{33-2n_f}\frac{1}{log(a+q^2/\Lambda^2_{QCD})}.\\
\end{aligned}
\end{equation}
where
$\alpha _s(q)$ is the running coupling constant, $\lambda$ is the string constant, $a$ and $\alpha$ are phenomenal parameters we introduce to avoid divergences when $q^2$ $\sim$ $\Lambda^2_{QCD}$ and $q^2$ $\sim$ $0$, $V_0$ is a constant in our model to fit the data.

\section*{Appendix B. Wavefunctions for different states}
In this section, we introduce the wavefunctions for different states.

\subsection*{B.1 Wave function for $^1S_0$}

The general form of $^1S_0$ state:

\begin{equation}\renewcommand\theequation{B.1.1}
\varphi(0^-)(\vec{q})=M\left[\frac{\slashed{P}}{M}f_1(\vec{q})+f_2(\vec{q})+\frac{\slashed{q}_\perp}{M}f_3(\vec{q})+\frac{\slashed{P}\slashed{q}_\perp}{M^2}f_4(\vec{q})\right]\gamma _5.
\end{equation}

Due to the constrains equations in full Salpeter equation, we have the condition $\varphi^{+-}_{0^-}$=$\varphi^{-+}_{0^-}$=0, Thus

\begin{equation}\renewcommand\theequation{B.1.2}
f_3(\vec{q})=\frac{f_2(\vec{q})M(\omega _2-\omega _1)}{m_1 \omega _2+m_2 \omega_1 }, f_4(\vec{q})=-\frac{f_1(\vec{q})M(\omega _2+\omega _1)}{m_1 \omega _2+m_2 \omega_1 }.
\end{equation}

Therefore, there are only two independent wavefunctions $f_1(\vec{q})$ and $f_2(\vec{q})$.
The relativistic positive wavefunction could be written as

\begin{equation}\renewcommand\theequation{B.1.3}
\varphi^{++}(^1S_0)(\vec{q})=a_1\left[\frac{a_2\slashed{P}}{M}+\frac{a_3\slashed{q}_\perp}{M}+\frac{a_4\slashed{q}_\perp\slashed{P}}{M^2}+1\right]\gamma ^5,
\end{equation}
where
\begin{equation*}
\begin{aligned}
&a_1=\frac{M}{2}\left(f_1(\vec{q})+f_2(\vec{q})\frac{\omega_1 + \omega_2 }{m_1+m_2}\right), \quad
a_2=\frac{m_1+m_2}{\omega_1 + \omega_2},\\
&a_3=-M\frac{\omega_1 - \omega_2}{m_1\omega_2 + m_2\omega_1}, \quad
a_4=M\frac{m_1+m_2}{m_1\omega_2 + m_2\omega_1}.
\end{aligned}
\end{equation*}

\subsection*{B.2 Wave function for $^1P_1$}
The general form of $^1P_1$ state:

\begin{equation}\renewcommand\theequation{B.2.1}
\varphi (^1P_1)(\vec{q}_f)=q_{f\perp}\cdot \varepsilon \left[g_1(\vec{q}_f)+g_2(\vec{q}_f)\frac{P_f}{M_f}+g_3(\vec{q}_f)\slashed{q}_{f\perp}+\frac{\slashed{P}_f\slashed{q}_{f\perp}}{M_f^2}g_4(\vec{q}_f)\right]\gamma _5.
\end{equation}

Constrains equations result in
\begin{equation}\renewcommand\theequation{B.2.2}
g_3(\vec{q}_f)=-\frac{\omega'_1 - \omega'_2}{m'_1\omega'_2 + m'_2\omega'_1}g_1(\vec{q}_f), \quad g_4(\vec{q}_f)=-\frac{(\omega'_1 + \omega'_2)M_f}{m'_1\omega'_2 + m'_2\omega'_1}g_2(\vec{q}_f).
\end{equation}

Thus the relativistic wavefunction is
\begin{equation}\renewcommand\theequation{B.2.3}
\begin{aligned}
\varphi^{++}(^1P_1)(\vec{q}_f)=&\frac{q_{f\perp} \cdot \varepsilon}{2} \left[g_1(\vec{q}_f)+\frac{\omega'_1 + \omega' _2}{m'_1+m'_2}g_2(\vec{q}_f)\right]\left[1+\frac{m'_1+m'_2}{\omega'_1 +\omega'_2}\frac{\slashed{P}_f}{M_f}-\frac{\omega'_1- \omega'_2 }{m'_1\omega'_2+m'_2\omega'_1}\slashed{q}_{f\perp} \right.\\
&\left. +\frac{m'_1+m'_2}{m'_1\omega '_2+m'_2\omega'_1}\frac{\slashed{q}_{f\perp} \slashed{P}_f}{M_f} \right]\gamma^5.
\end{aligned}
\end{equation}

The Dirac conjugate form is:
\begin{equation}\renewcommand\theequation{B.2.4}
\overline{\varphi}^{++}(^1P_1)(\vec{q}_f)=-\frac{\varepsilon \cdot q_{f\perp}}{2} a_5 \gamma ^5 \left(1+a_7\frac{\slashed{P}_f}{M}_f+a_8\slashed{q}_{f\perp}+a_9\frac{\slashed{P}_f\slashed{q}_{f\perp}}{M}_f\right),
\end{equation}
where
\begin{equation*}
\begin{aligned}
&a_5=g_1(\vec{q}_f)+g_2(\vec{q}_f)\frac{w'_1+w'_2}{m'_1+m'_2}, \quad
a_7=\frac{m'_1+m'_2}{w'_1+w'_2},\\
&a_8=-\frac{w'_1+w'_2}{m'_1w'_2+m'_2w'_1}, \qquad \quad
a_9=\frac{m'_1+m'_2}{m'_1w'_2+m'_2w'_1}.
\end{aligned}
\end{equation*}

\subsection*{B.3 Wave function for $^3P_1$}
In the same way, we have the wavefunction of $^3P_1$ state:
\begin{equation}\renewcommand\theequation{B.3.1}
\begin{aligned}
\varphi^{++}(^3P_1)(\vec{q}_f)=&\frac{i}{2M_f}\left[h_1(\vec{q}_f)+\frac{\omega' _1+\omega' _2}{m'_1+m'_2}h_2(\vec{q}_f)\right]\left[1+\frac{m'_1+m'_2}{\omega '_1+\omega '_2}\frac{\slashed{P}_f}{M}_f-\frac{\omega' _1-\omega' _2}{m'_1\omega '_2+m'_2\omega '_1}\slashed{q}_{f\perp}\right.\\
&\left.+\frac{m'_1+m'_2}{m'_1\omega '_2+m'_2\omega'_1}\frac{\slashed{q}_{f\perp} \slashed{P}_f}{M_f}i\epsilon _{\nu \lambda \rho \sigma }\gamma ^\nu P^\lambda_f q^\rho_{f\perp} \epsilon ^\sigma \right],
\end{aligned}
\end{equation}
and it's Dirac conjugate
\begin{equation}\renewcommand\theequation{B.3.2}
\overline{\varphi}^{++}(^3P_1)(\vec{q}_f)=-\frac{i}{2M_f}a_6\epsilon _{\nu \lambda \rho \sigma} \gamma ^\nu P^\lambda_f q_{f\perp}^\rho \varepsilon ^\sigma \left(1+a_7\frac{\slashed{P}_f}{M_f}+a_8 \slashed{q}_{f\perp}+a_9\frac{\slashed{P}_f\slashed{q}_{f\perp}}{M_f}\right),
\end{equation}
where
\begin{equation*}
\begin{aligned}
&a_6=h_1(\vec{q}_f)+h_2(\vec{q}_f)\frac{w'_1+w'_2}{m'_1+m'_2}.
\end{aligned}
\end{equation*}


\section*{Appendix C. The form factor}
In this section, we present the form factors in semi-leptonic decay of $B^0_{s}$ into $D_s(2P)$ state. For the process of $D_J(2P)$, the form factors are the same.

\begin{eqnarray*}
t_1=&&\frac{a_1a_5}{2M^2M_{f1}^2}(2\alpha ^2E_{f1}(E_{f1}^2a_9M+a_2E_{f1}a_8MM_{f1}+a_4E_{f1}a_9 P_{f1}\cdot q+a_3a_8M_{f1}P_{f1}\cdot q)+2\alpha E_{f1}\\
&&\times (-MM_{f1}+2E_{f1}M(E_{f1}a_9+a_2a_8M_{f1})X+
a_4E_{f1}(M_{f1}+2a_9P_{f1}\cdot q)X+a_3(a_7(P_{f1}\cdot q+E_{f1}^2X)\\
&&+a_8M_{f1}(q_\perp^2 +P_{f1}\cdot q)))+
E_{f1}(2(-MM_{f1}+a_3a_7P_{f1}\cdot q+a_3a_8M_{f1}q_\perp^2)X-
E_{f1}(a_3E_{f1}a_7\\
&&+E_{f1}a_9M+a_4M_{f1}+a_2a_8MM_{f1}+a_4a_9P_{f1}\cdot q)q_\perp^2 Y+
(a_3E_{f1}a_7+E_{f1}a_9M+a_4M_{f1}\\
&&+a_2a_8MM_{f1}+a_4a_9P_{f1}\cdot q)q_\perp^2 Z),\\
t_2=&&\frac{a_1a_5}{2M^2M_{f1}^2}E_{f1}(-2\alpha M(\alpha E_{f1}a_9+a_2(a_7+\alpha a_8M_{f1}))+2\alpha a_4a_9q_\perp^2-
2(a_2a_7M+\alpha (a_3E_{f1}a_7\\
&&+2E_{f1}a_9M+a_4M_{f1}+2a_2a_8MM_{f1}+a_4a_9P_{f1}\cdot q)-a_4a_9 q_\perp^2)X+
(a_3E_{f1}a_7+E_{f1}a_9M+a_4M_{f1}\\
&&+a_2a_8MM_{f1}+a_4a_9P_{f1}\cdot q)q_\perp^2 Y),\\
t_3=&&-\frac{a_1a_5(a_3E_{f1}a_7+E_{f1}a_9M+a_4M_{f1}+a_2a_8MM_{f1}+a_4a_9P_{f1}\cdot q)q_\perp^2 Z}{2MM_{f1}},\\
t_4=&&\frac{ia_1a_5(a_9(\alpha a_4E_{f1}+M)+a_3(a_7+\alpha a_8M_{f1}))q_\perp^2Z}{2M^2M_{f1}},\\
t_5=&&\frac{a_1a_6}{2M^2M_{f2}^2}(2\alpha ^2E_{f2}(a_2E_{f2}a_9MM_{f2}^2+a_8MM_{f2}^3+CE_{f2}^2a_9P_{f1}\cdot q+a_4E_{f2}a_8M_{f2}P_{f1}\cdot q +E_{f2}\\
&&(E_{f2}-M_{f2})(E_{f2}+M_{f2})(a_3E_{f2}a_9+a_4a_8M_{f2})X)+E_{f2}(2(a_7MM_{f2}^2+a_8MM_{f2}P_{f1}\cdot q\\
&&-a_3P_{f1}\cdot q(M_{f2}+a_9P_{f1}\cdot q)+a_3a_9M_{f2}^2 q_\perp^2)X
+E_{f2}(M_{f2}(a_3E_{f2}+a_4a_7M_{f2}-M(E_{f2}a_8+a_2a_9M_{f2}))\\
&&+a_3E_{f2}a_9P_{f1}\cdot q) q_\perp^2 Y)-(M_{f2}(a_3E_{f2}+a_4a_7M_{f2}-M(E_{f2}a_8+a_2a_9M_{f2}))+a_3E_{f2}a_9 P_{f1} \cdot q)\\
&&q_\perp^2 Z+
\alpha (E_{f2}(2(-a_3M_{f2}P_{f1}\cdot q+a_8MM_{f2}P_{f1}\cdot q-a_3a_9(P_{f1}\cdot q)^2+a_3a_9M_{f2}^2q_\perp^2\\
&&+M_{f2}(-a_3E_{f2}^2+E_{f2}^2a_8M+2a_2E_{f2}a_9MM_{f2}+a_8MM_{f2}^2+a_4E_{f2}a_8P_{f1}\cdot q)X\\
&&+a_7M_{f2}^2(M-a_4E_{f2}X))
+E_{f2}(-E_{f2}+M_{f2})(E_f2+M_{f2})(a_3E_{f2}a_9+a_4a_8M_{f2})q_\perp^2 Y)\\
&&+(E_{f2}-M_{f2})(E_{f2}+M_{f2})(a_3E_{f2}a_9+a_4a_8M_{f2})q_\perp^2 Z)),\\
t_6=&&\frac{a_1a_6}{2M^2M_{f2}^2}(-2\alpha ^2E_{f2}^2M(a_2E_{f2}a_9+a_8M_{f2})-2a_3M_{f2}q_\perp^2+2M(a_7P_{f2}\cdot q+a_8M_{f2}q_\perp^2)\\
&&-2(a_7(MM_{f2}^2+a_4E_{f2}P_{f2} \cdot q)-P_{f2} \cdot q(a_2E_{f2}a_9M+a_3M_{f2}-a_8MM_{f2}+a_3a_9P_{f2}\cdot q)\\
&&+a_3a_9M_{f2}^2 q_\perp^2)X+2\alpha (E_{f2}^3(a_4a_7-a_2a_9M)X+a_8MM_{f2}(P_{f2} \cdot q-M_{f2}^2X)-E_{f2}^2(a_7M+(-a_3M_{f2}\\
&&+a_8MM_{f2}+a_3a_9P_{f2} \cdot q)X)+E_{f2}(a_2a_9M(P_{f2} \cdot q-M_{f2}^2X)+a_4a_8M_{f2}(q_\perp^2 - 2 P_{f2} \cdot qX)))\\
&&+\alpha E_{f2}(E_{f2}-M_{f2})(E_{f2}+M_{f2})(a_3E_{f2}a_9+a_4a_8M_{f2})q_\perp^2 Y-
E_{f2}(M_{f2}(a_3E_{f2}+a_4FM_{f2}\\
&&-M(E_{f2}a_8+a_2a_9M_{f2}))+a_3E_{f2}a_9P_{f2} \cdot q)q_\perp^2Y),\\
t_7=&&\frac{a_1a_6}{2MM_{f2}^2}(2\alpha ^2E_{f2}M(E_{f2}-M_{f2})(E_{f2}+M_{f2})(a_2E_{f2}a_9+a_8M_{f2})-(a_4a_7-a_2a_9M)(2(P_{f2} \cdot q)^2\\
&&-M_{f2}^2 q_\perp^2(2+Z))+E_{f2}(-2a_7MP_{f2} \cdot q+ q_\perp^2 (-a_8MM_{f2}(2+Z)+a_3(a_9P_{f2} \cdot q Z\\
&&+M_{f2}(2+Z))))+\alpha(E_{f2}^2(2a_7M-a_3a_9 q_\perp^2 Z)+E_{f2} M_{f2}(-2a_7MM_{f2}+2a_3 P_{f2} \cdot q- 4a_8MP_{f2} \cdot q
\end{eqnarray*}
\begin{eqnarray*}
&&+a_3a_9M_{f2} q_\perp^2 Z)+a_4a_8M_{f2}(-2(P_{f2} \cdot q)^2+M_{f2}^2 q_\perp^2(2+Z))+E_{f2}^2(-4a_2a_9MP_{f2} \cdot q\\
&&+a_4(2a_7P_{f2} \cdot q-a_8M_{f2} q_\perp^2(2+Z))))), \\
t_8=&&-\frac{a_1a_6}{2M^2M_{f2}^2}i(2E_{f2}(a_4a_7P_{f2} \cdot q+M_{f2}(a_2M+a_4a_8(\alpha P_{f2} \cdot q + q_\perp^2))+E_{f2}(a_7M+a_3a_9(\alpha P_{f2} \cdot q \\
&&+ q_\perp^2)))(\alpha +X)+(a_8MM_{f2}+\alpha E_{f2}(a_3E_{f2}a_9+a_4a_8M_{f2})-a_3(M_{f2}+a_9 P_{f2} \cdot q)) q_\perp^2 Z),
\end{eqnarray*}
where $E_{f1}$ and $E_{f2}$ are the energies of $^1P_1$ and $^3P_1$ states, $M_{f1}$ and $M_{f2}$ are the masses of $^1P_1$ and $^3P_1$ states. $X=\frac{q\cos \theta}{|\vec{P}_f|}$,$Y=\frac{-1+3\cos^2\theta}{|\vec{P}_f|}$,$Z=-1+\cos^2\theta$.



\end{document}